# DRP-VEM: Drug repositioning prediction using voting ensemble

Zahra Ghorbanali, Fatemeh Zare-Mirakabad*, Bahram Mohammadpour

- **Zahra Ghorbanali**, Ph.D. student of computer science, Amirkabir University of Technology, Email:z_ghorbanali@aut.ac.ir
- **Fatemeh Zare-Mirakabad**, Assistant Professor of Computer Science, Amirkabir University of Technology, Email:f.zare@aut.ac.ir (*Corresponding Author)
- **Bahram Mohammadpour,** Master student of computer science, Amirkabir University of Technology, Email: bahram.mp@gmail.com

## Abstract

Traditional drug discovery methods are costly and time-consuming. Drug repositioning (DR) is a common strategy to overcome these issues. Recently, machine learning methods have been used extensively in DR problem. The performance of these methods depends on the features, representations and training dataset. In this problem, feature sets include many redundant features, which have a negative effect on the performance of methods. Moreover, selecting an appropriate training set is influential in the rise of machine learning method accuracy. However, in this problem, we face two obstacles to find the proper training set. First, most methods employ known and unknown drug-disease pairs as positive and negative sets, respectively. While the number of known pairs is much less than unknowns, it leads to machine learning performance error because of biasing to the majority group. Second, the absence of a drug-disease association means this association has not been approved experimentally and may be changed. In this paper, DRP-VEM framework is proposed to overcome the challenges. We assess DRP-VEM based on different parameters: disease and drug feature representations, classification methods, and voting ensemble training approaches. DRP-VEM is evaluated using heterogenous evaluation criteria. Moreover, we compare DRP-VEM using the best combination of parameters with DisDrugPred.

# Introduction

Despite the growth of technology and its role in diagnosing diseases, transforming these successes and benefits to medical treatment is not fast enough [1]. Traditional drug discovery is a time-consuming and costly method. According to recent studies, the process of discovering a new druggable component, passing the test phase steps, and bringing it to markets takes more than ten years and from $314 million to $2.8 billion [2]. Drug repositioning or drug repurposing uses an approved drug for a new indication outside its first treatment purpose. A historical example of drug repositioning is Sildenafil. Researchers developed Sildenafil for treating hypertension, but today it is used to cure erectile dysfunction and is known as Viagra [1]. This method also can be used for treating new diseases. For example, researchers repurposed existing antiviral drugs such as Baloxavir, Azvudine, and Darunavir to treat coronavirus disease during the Covid-19 pandemic [3].

Firstly, some physicians discovered drug repositioning opportunistically. Although using retrospective clinical experiences is useful, physicians have to check a wide range of drugs. Despite the increased cost and risk of failure, finding an alternative drug to treat a specific disease would become time-consuming. As these methods have not involved a systematic approach, nowadays, due to the growth of computational methods and their application in various studies and the expansion of available data of drugs and diseases, researchers prefer to apply computational methods for solving drug repositioning problem. We divide these computational methods into three main groups: drug-based, disease-based, and hybrid.

Owing to the availability of drug information, more researchers focus on drug-based techniques. Ozsoy et al. combined three main drug features: chemical structure, protein interaction, and side effects. Pareto dominance technique was applied to find the neighbors of a drug. Then, they used a collaborative filtering recommendation system to find the probability of association between drug-disease pairs [4]. Zeng et al. proposed the DeepDR method, which performed random walk to represent drug feature networks and then combined them using an autoencoder. Finally, a variational autoencoder was applied to estimate the drug-disease association probabilities [5]. Chen et al. collected three features of the drug: chemical structure, targets, and side effects. After calculating the similarities of drugs, a fusion method was developed for merging the similarities to predict the probabilities of drug-disease associations [6].

Despite the importance of disease-related data in the drug repositioning problem, researchers have not widely studied disease-based methods due to a lack of information. Therefore, these methods have focused on a specific disease or therapeutic domain [7]. Chiang and Butte calculated the similarities of the diseases by counting shared therapies. Then, a " guilt by association " approach was applied to consider disease similarities. By using these similarities, they found new drug-disease association pairs [8].

In hybrid methods, researchers combine both drug and disease data to obtain the chance of drug-disease association pair. Moridi et al. presented a pipeline that efficiently represents drug and disease features using the deep learning method. They proposed a non-linear approach to find the drug-disease candidates [9]. Xuan et al. presented DisDrugPred by integrating drug similarities, disease similarities and known drug-disease associations using non-negative matrix factorization technique to calculate the association probability of drugs and diseases [10]. Lue et al. proposed an approach named RWHND to reconstruct a heterogeneous network by combining drugs, drug targets, diseases and disease genes data. Then, a random walk model was developed to candidate pharmaceutical treatment for a disease [11].

Although researchers have done great studies on the drug repositioning (DR) problem, challenges still need to be addressed. In the following, we review these challenges and our idea to overcome them:

- The previous studies focused on drug-based methods mostly and less on the hybrid. In addition, these studies used different drug features and tried to combine all of them. However, they have not focused on which feature has a significant role in detecting drug-disease association pairs or the appropriate representation. In this paper, we aim to ascertain if using all features in solving

- DR problems is necessary or causes redundancy. Moreover, we find significant feature which improves the accuracy of machine learning methods in addressing DR problem. In addition, we assess which data representations have better performance than others.
- In the literature, different machine learning techniques can be found to solve DR problem. This article shows that if the selected representation and combination of features are defined appropriately, the effect of classification methods on predicted results for the DR problem differs slightly.
- Most approaches consider known drug-disease pairs as positive and all unknown pairs as negative sets. Nevertheless, finding the proper training set faces two challenges. In the first one, while the number of known pairs is much less than unknown ones, it leads machine learning biasing to the leading group, so the method's performance is flawed [12]. The second one, the lack of a drug and disease association as a negative set, has not been assessed clinically. This study introduces a new algorithm to make a training set called voting ensemble training approach to overcome this issue.

To show that the selected drug feature, the chosen feature representation of drugs and disease, the elected machine learning method and the voting ensemble training approach are suitable, we compare our framework with the DisDrugPred [10].

The rest of this article is constructed as follows: the "Methods" section presents the description of DR problem, data and our framework called DRP-VEM. The "Results and Discussion" section includes the assessment of DRP-VEM and comparison results with DisDrugPred, and finally, the "Conclusion" shows the future point of DR problem.

# Methods

This article aims to:

- find which feature presentations are appropriate to depict the drugs and diseases in DR problem,
- analyze which feature has more impact on solving DR problem,
- assess if all drug features are necessary or cause redundancy for predicting drug-disease associations,
- select which classification method shows more accuracy for DR problem,
- propose a voting ensemble training approach to overcome challenges about using unknown drug-disease pairs as the negative set and unbalanced data in facing DR problem.

In the following section, first, we define DR problem. Ensuing, we introduce our datasets. Next, we illustrate the data representations and training approach. Finally, we propose and explain our method.

## Drug repositioning problem

A disease set and a drug set are shown by $\rho = \{P_1, P_2, ..., P_n\}$ and $\varphi = \{R_1, R_2, ..., R_m\}$, respectively, where, $n$ and $m$ are the numbers of diseases and drugs. In the mathematical definition of DR problem, our primary goal is to find the existence of a therapeutic association between disease $P \in \rho$ and drug $R \in \varphi$. If the model predicts $<P, R>$ has a therapeutic association, the output is one and otherwise zero. In DR problem, the following data is given as input:

- The features of the disease $P$,
- The features of the drug $R$,
- The set of known drug-disease association pairs,
  $A = \{(P_i, R_j) | P_i \in \rho, R_j \in \varphi$

  there is known assciation between disease $P_i$ and drug $R_j\}$.

## Data sources

It is necessary to collect some known drug-disease associations and select some features for drugs and diseases. Therefore, we use four drug features: target, domain, side effect, and chemical structure. Also, we apply semantic similarity as a disease feature. In the following, the databases used to extract data are introduced:

- Drug-Disease association: we choose the "repoDB" database [13] to collect known drug-disease association pairs.
- Drug features: We retrieve drug names, identification, and target from "DrugBank" [14]. The target domains of drugs are extracted from "Uniprot" [15]. We derive the information on side effects from "SIDER4.1" [16]. Finally, the chemical structures of drugs are collected from "Pubchem" [17].
- Disease Feature: We extract the disease similarity from "DincRNA" [18] based on Wang's method [19].

The list of diseases is limited based on the DincRNA database with size of 158. In addition, we select 413 drugs where all features are available. Table 1 shows the number of extracted data from databases. Here, there are 1506 target components for drugs. Also, there are 1070 domain, 5734 side effect and 881 chemical structure components.

Table 1: The size of databases

| Data | Size |
| --- | --- |
| Targets (T) | 1506 |
| Domains (D) | 1070 |
| Side effects (S) | 5734 |
| Chemical structure (C) | 881 |

## Data representation

In this subsection, we introduce how to present each data for feeding into the framework.

- **Drug feature representation**

We define $\mathcal{F}_R^F = \{\xi_1, \xi_2, ..., \xi_{l_F}\}$ as a set of feature components for $F \in \{T, D, S, C\}$, where $\xi_i$ shows $i_{th}$ component of feature $F$, and $l_F$ represents the number of feature components. Each feature $T, D, S,$ and $C$, shows target, domain, side effect, and chemical structure, respectively. In the following, we introduce two types of drug representation named binary (B) and cosine similarity (C), as follows:

✓ The binary vector $B_R^F$ with length $l_F$ is defined for drug $R$ based on feature $F$, as:

$$\forall 1 \leq k \leq l_F : B_R^F[k] = \begin{cases} 1 & \text{component } \xi_k \text{ is related to drug } R, \\ 0 & \text{otherwise.} \end{cases} \quad (1)$$

The binary vectors $B_R^T$, $B_R^D$, $B_R^S$ and $B_R^C$ are computed for target, domain, side effect and chemical structure features based on (1), respectively. In addition, we define a binary vector, $B_R^G$, based on concatenation of all drug features as below:

$$B_R^G = B_R^T . B_R^D . B_R^S . B_R^C \quad (2)$$

✓ The cosine similarity vector $C_R^F$ with length $m$ is presented for drug $R$ based on feature $F$ as follows:

$$\forall 1 \leq k \leq m, R_k \in \varphi = \{R_1, R_2, ..., R_m\}$$
$$C_R^F[k] = \text{cosine similarity between } B_R^F \text{ and } B_{R_k}^F. \quad (3)$$

The cosine similarity vectors $C_R^T$, $C_R^D$, $C_R^S$ and $C_R^C$ are computed for target, domain, side effect and chemical structure features based on (3), respectively. Also, we define two cosine similarity vectors, $C_R^G$ and $C_R^N$, based on the concatenation and normalizing of all drug features, as follows, respectively:

$$C_R^G = C_R^T . C_R^D . C_R^S . C_R^C, \quad C_R^N = Norm(C_R^T + C_R^D + C_R^S + C_R^C) \quad (4)$$

Thus, we define eleven different drug feature representations as bellow:

$$\mathbb{U} = \{B^T, B^D, B^S, B^C, B^G, C^T, C^D, C^S, C^C, C^G, C^N\}.$$

- **Disease feature representation**

Here, we define two types of disease representation called Wang vector (W) and one-hot vector (O) as below, $\mathbb{V} = \{O, W\}$:

✓ Wang vector $W_P$ with size $n$ is defined for disease $P$ based on Wang's similarity function [19], where:

$$\forall 1 \leq k \leq n, P_k \in \rho = \{P_1, P_2, ..., P_n\}$$
$$W_P[k] = Wang's\ similarity\ between\ disease\ P\ and\ P_k. \quad (5)$$

✓ The one-hot vector $O_P$ with length $n$ is defined for disease $P$ as follows:

$$\forall 1 \leq k \leq n, P_k \in \rho = \{P_1, P_2, ..., P_n\}$$
$$O_P[k] = \begin{cases} 1 & P = P_k, \\ 0 & otherwise. \end{cases} \quad (6)$$

## The voting ensemble training approach

This subsection suggests an approach to make training and test sets for a classification model. As a common approach to make training and test sets for DR problem, known and unknown drug-disease association pairs are considered as positive (A) and negative (B) sets, respectively, where:

$$A = \{(P_i, R_j) | P_i \in \rho, R_j \in \varphi,$$
$$there\ is\ a\ known\ association\ btween\ disease\ P_i\ and\ drug\ R_j\},$$

$$B = \{(P_i, R_j) | P_i \in \rho, R_j \in \varphi,$$
$$there\ is\ no\ known\ association\ btween\ disease\ P_i\ and\ drug\ R_j\}.$$

However, this approach is not appropriate because of two main challenges:

✓ The number of known pairs is much less than unknown ones $(A \ll B)$. It leads binary classifier biasing to the majority group, so the method's performance is flawed [12].
✓ The lack of a drug and disease association means the association of this pair has not been assessed clinically yet, not that the pair will never be associated.

To overcome the first challenge, we apply an under-sampling approach by randomly selecting unknown association pairs with size k-times of known association pairs. We called this selecting approach for making training set as one-to-k distribution.

To address the second challenge, we cluster unknown pairs according to the one-to-k distribution approach for constructing negative training sets in which their intersection set is empty and the union set equals the whole. Assume that the number of these clusters is $p_k$. So, the model is trained on $p_k$ negative datasets. For each test sample, we vote the response of the trained models to predict association. The details of making training and test sets based on voting ensemble method are available as follows:

1. Set A, set B and an integer number k are given as inputs.
2. Set $P_{test}$ is the positive test set and includes $1/10$ randomly chosen samples from set $A$.
3. Set $N_{test}$ is the negative test set and includes $k * |P_{test}|$ randomly chosen samples from set $B$.
4. Set $P_{train} = A - P_{test}$ is the positive training set.
5. Set $N_{train_i}$ is the $ith$ negative training set based on a one-to-k distribution approach and includes $k * |P_{train}|$ randomly chosen samples from set $B - N_{test} - (\bigcup_{j=1}^{i-1} N_{train_j})$. Assume that we can define $p_k$ different negative training sets where $\bigcup_{i=1}^{p_k} N_{train_i} = B - N_{test}$.
6. Set $\varepsilon = P_{test} \cup N_{test}$ is considered as the test set

7. Set $\tau_i = P_{train} \cup N_{train_i}$ is known as $ith$ the training set.
8. Set $T_k = \{\tau_i \mid 1 \leq i \leq p_k\}$ is defined as the voting ensemble training set.

For each k=1,2,3,5, we define $T_k$ as a voting ensemble training set, including $p_k$ sets for training a classifier. Each $\tau_i \in T_k$ is fed to a classifier as the training set. Therefore, the classifier is trained $p_k$ times. For each sample from the test set, such as $x \in \varepsilon$, each trained model predicts the association between disease and drug. Finally, we vote on $p_k$ predicted results (see figure 1).

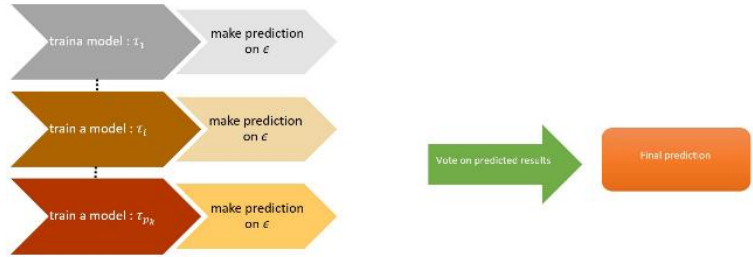

Fig. 1. The voting ensemble trainig approach.

## Method

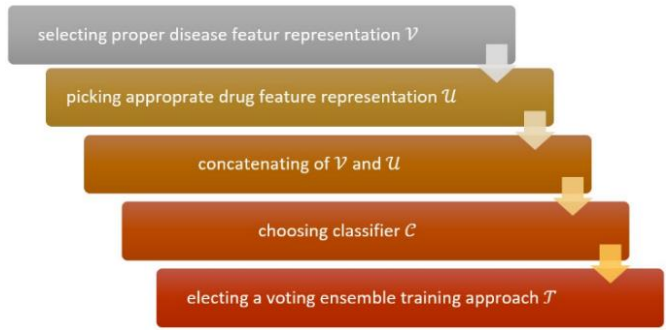

Fig. 2. The main step of DRP-VEM.

The main steps of the DRP-VEM, $model = <\mathcal{V},\mathcal{U},\mathcal{C},\mathcal{T}>$, are available in figure 2 and the details are as follows:

1. Selecting a proper disease representation, $\mathcal{V} \in \mathbb{V} = \{O, W\}$.
2. Picking an appropriate drug representation, $\mathcal{U} \in \mathbb{U} = \{B^T, B^D, B^S, B^C, B^G, C^T, C^D, C^S, C^C, C^G, C^N\}$
3. Concatenating $\mathcal{V}$ and $\mathcal{U}$, $\mathcal{P} = \mathcal{V}.\mathcal{U}$, for feeding into a classifier.
4. Choosing decision tree ($DT$), random forest ($RF$), and complement naïve bayes ($CNB$) classifiers named $\mathcal{C} \in \mathbb{C} = \{DT, RF, CNB\}$.
5. Electing a voting ensemble training approach called $\mathcal{T} \in \mathbb{T} = \{T_1, T_2, T_3, T_5\}$.

We assess the different combinations of parameters to find the best model based on our framework.

## Results and discussion

This section evaluates $model = <\mathcal{V},\mathcal{U},\mathcal{C},\mathcal{T}>$ (see figure 2) based on different parameters, disease feature representation $\mathcal{V} \in \mathbb{V} = \{O, W\}$, drug feature representation $\mathcal{U} \in \mathbb{U} = \{B^T, B^D, B^S, B^C, B^G, C^T, C^D, C^S, C^C, C^G, C^N\}$, classifier methods $\mathcal{C} \in \mathbb{C} = \{DT, RF, CNB\}$, and voting ensemble training approach $\mathcal{T} \in \mathbb{T} = \{T_1, T_2, T_3, T_5\}$. Therefore, we train 264 models.

Here, we introduce the evaluation criteria, analyze our results, and compare our method with DisDrugPred [10].

Evaluation Criteria

We choose four different evaluation criteria, including accuracy (ACC), area under receiver operating characteristic curve (AUC), area under the precision-recall curve (AUC-PR), and weighted average score (WAS), to evaluate every $model^z$, where $1 \leq z \leq 264$.

ACC shows the rate of correct prediction to all predictions as below:

$$ACC = \frac{TP + TN}{TP + TN + FP + FN}, \quad (7)$$

where the definition of true positive (TP), false positive (FP), true negative (TN) and false negative (FN) is available in table 2.

Table2: Definition of TP, FP, TN, and FN

| Prediction | Definition |
|---|---|
| **True Positive (TP)** | the number of known drug-disease association pairs predicted correctly by the model |
| **False Positive (FP)** | the number of unknown drug-disease association pairs predicted wrongly by the model |
| **True Negative (TN)** | the number of unknown drug-disease association pairs predicted correctly by the model |
| **False Negative (FN)** | the number of known drug-disease association pairs predicted wrongly by the model |

AUC [20] is the area under the receiver operating characteristic curve, which uses different ranking cutoffs and curves the true positive rate (*TPR*) and false positive rate (*FPR*) where,

$$TPR = \frac{TP}{TP + FN}, \quad FPR = \frac{FP}{FP + TN}. \quad (8)$$

AUC-PR [21] shows the area under a precision-recall curve, a plot of precision and recall, where:

$$precision = \frac{TP}{TP + FP}, \quad recall = \frac{TP}{TP + FN}. \quad (9)$$

WAS is calculated based on integrating ACC, AUC, and AUC-PR as below:

$$WAS = \frac{2*AUC + 2*AUC-PR + ACC}{5}. \quad (10)$$

We define the set of evaluation criteria as $\mathbb{E} = \{ACC, AUC, AUC\text{-}PR, WAS\}$. For each $e \in \mathbb{E}$ and $z$, $1 \leq z \leq 264$, the evaluation of $model^z = <\mathcal{V}^z, \mathcal{U}^z, \mathcal{C}^z, \mathcal{T}^z>$ on the test set is shown by $e(model^z)$.

Assessment of DRP-VEM based on different parameters

In the following, we assess our framework, DRP-VEM, based on the selected characteristics of each parameter.

- **Assessment of disease feature representation**

We defined W and O's disease feature representations based on the Wang vector (5) and the one-hot vector (6). The performance of each $\mathcal{V} \in \mathbb{V} = \{O, W\}$ according to every evaluation criterion, $e \in \mathbb{E}$, is calculated based on (11):

$$avg(\mathcal{V}^z = \mathcal{V}) = \frac{1}{|\mathbb{T}|.|\mathbb{C}|.|\mathbb{U}|} \sum_{\mathcal{T}^z \in \mathbb{T}} \sum_{\mathcal{C}^z \in \mathbb{C}} \sum_{\mathcal{U}^z \in \mathbb{U}} e(model^z). \quad (11)$$

Table 3 shows the value of $avg(\mathcal{V}^z = \mathcal{V})$ for each $\mathcal{V} \in \mathbb{V}$. Although the scores are close, the Wang vector performs better than the one-hot vector. So, we can apply this representation for diseases.

*table3:Evaluation criteria on disease representation.*

| Disease representation ($\mathcal{V}$) | $e = ACC$ | $e = AUC$ | $e = AUC - PR$ | $e = WAS$ |
|---|---|---|---|---|
| W | %77.2 | **%69.8** | **%51.8** | **%64** |
| O | **%77.4** | %69.4 | %50.3 | %63 |

- **Assessment of drug feature selection and representation**

We determined eleven drug feature representations from set $\mathcal{U} \in \mathbb{U} = \{B^T, B^D, B^S, B^C, B^G, C^T, C^D, C^S, C^C, C^G, C^N\}$.

The performance of each $\mathcal{U} \in \mathbb{U}$ for every evaluation criterion $e \in \mathbb{E}$ is measured according to (12):

$$avg(\mathcal{U}^z = \mathcal{U}) = \frac{1}{|\mathbb{T}|.|\mathbb{C}|.|\mathbb{V}|} \sum_{\mathcal{T}^z \in \mathbb{T}} \sum_{\mathcal{C}^z \in \mathbb{C}} \sum_{\mathcal{V}^z \in \mathbb{V}} e(model^z). \quad (12)$$

The evaluation scores of every drug feature representation are shown in table 4. Applying target cosine similarity vector ($C^T$) or domain cosine similarity vector ($C^D$) with a slight difference ($\sim 0.8\%$) has better performances than the other drug feature representations. We can infer that the target of a drug has a significant effect on predicting the association of a drug-disease pair. Meanwhile, the domain of a target is a region of the protein's polypeptide chain that is self-stabilizing and folds independently from the rest [22]. As their scores are very close, applying one of these features is required. Assessing extracted results of table 4 shows implying target or domain feature is essential. While the combination of features (see $B^G$, $C^G$ and $C^N$ in table 4) reduces the model's performance.

*Table4 :Evaluation criteria drug representation*

| Drug representation ($\mathcal{U}$) | $e = ACC$ | $e = AUC$ | $e = AUC - PR$ | $e = WAS$ |
|---|---|---|---|---|
| $B^T$ | %80.0 | %72.7 | %55.1 | %67.1 |
| $B^D$ | %80.7 | %74.2 | %54.2 | %67.5 |
| $B^S$ | %79.4 | %70.0 | %53.0 | %65.1 |
| $B^C$ | %76.2 | %68.1 | %49.6 | %62.3 |
| $B^G$ | %77.5 | %67.1 | %48.7 | %61.9 |
| $C^T$ | **%81.0** | %74.2 | **%57.3** | **%68.8** |
| $C^D$ | %78.6 | **%74.3** | %56.5 | %68.0 |
| $C^S$ | %69.7 | %61.8 | %40.0 | %54.6 |
| $C^C$ | %76.5 | %70.6 | %52.4 | %64.5 |
| $C^G$ | %76.9 | %68.9 | %51.1 | %63.4 |
| $C^N$ | %73.8 | %63.7 | %43.7 | %57.7 |

- **Assessment of classification method**

We examined three state-of-the-art classifiers from set $\mathbb{C} = \{DT, RF, CNB\}$ where $DT$, $RF$, and $CNB$ present decision tree, random forest, and complement naïve bayes, respectively. The performance of each $\mathcal{C} \in \mathbb{C}$ for every evaluation criterion $e \in \mathbb{E}$ is calculated according to (13). The results are available in table 5.

$$avg(\mathcal{C}^z = \mathcal{C}) = \frac{1}{|\mathbb{T}|.|\mathbb{U}|.|\mathbb{V}|} \sum_{\mathcal{T}^z \in \mathbb{T}} \sum_{\mathcal{U}^z \in \mathbb{U}} \sum_{\mathcal{V}^z \in \mathbb{V}} e(model^z).$$
(13)

*Table 5: Evaluation criteria on classification method.*

| Classifier ($\mathcal{C}$) | e = Acc | e = AUC | e = AUC − PR | e = WAS |
|---|---|---|---|---|
| DT | %84.0 | %77.0 | %61.6 | %72.2 |
| RF | %80.3 | %70.0 | %52.8 | %65.2 |
| CNB | %67.6 | %61.9 | %38.7 | %53.8 |

According to table 5, DT distinguishes our dataset significantly better than two other models, and RF performance is more reliable than CNB.

- **Assessment of voting ensemble training approach**

We displayed four voting ensemble training sets, $\mathbb{T} = \{T_1, T_2, T_3, T_5\}$. Model $model^z$ is trained based on $\mathcal{T} \in \mathbb{T}$ and then prediction for test data is made according to figure 1. Table 6 illustrates every evaluation criterion $e \in \mathbb{E}$ for each $\mathcal{T} \in \mathbb{T}$ according to (14). The most accurate performance belongs to one-to-one distribution, $T_1$, where the number of positives and negatives are equal.

$$avg(\mathcal{T}^z = \mathcal{T}) = \frac{1}{|\mathbb{C}|.|\mathbb{U}|.|\mathbb{V}|} \sum_{\mathcal{C}^z \in \mathbb{C}} \sum_{\mathcal{U}^z \in \mathbb{U}} \sum_{\mathcal{V}^z \in \mathbb{V}} e(model^z).$$
(14)

*Table 6: Evaluation criteria on the training approach*

| Training approach ($\mathcal{T}$) | e = Acc | e = AUC | e = AUC − PR | e = WAS |
|---|---|---|---|---|
| $T_1$ | %71.1 | **%71.3** | **%66.2** | **%69.2** |
| $T_2$ | %75.2 | %69.6 | %52.7 | %64.0 |
| $T_3$ | %79.9 | %70.0 | %48.3 | %63.3 |
| $T_5$ | **%83.0** | %67.6 | %36.9 | %58.4 |

- **Assessment concatenation between different dug and disease features**

We fed the different combinations of the drug ($\mathcal{U} \in \mathbb{U}$) and the disease representation ($\mathcal{V} \in \mathbb{V}$) to $model^z$. We examine possible different combinations of them according to (15).

$$avg(\mathcal{V}^z = \mathcal{V}, \mathcal{U}^z = \mathcal{U}) = \frac{1}{|\mathbb{T}|.|\mathbb{C}|} \sum_{\mathcal{T} \in \mathbb{T}} \sum_{\mathcal{C} \in \mathbb{C}} e(model^z)$$
(15)

Table 7 depicts the performance of each combination. As we expected, the combination of the Wang vector (W) and target cosine similarity ($C^T$) has better results than other ones. The binary vector combined with the one-hot vector ($B \cdot O$) and cosine similarity vector with the Wang vector ($C \cdot W$) performs better in most cases. It seems to be because $B$ and $O$ are both discrete representations, as well $C$ and $W$ are continues one. Therefore, their combinations work more accurately.

Table7: Evaluation criteria on combination of disease and drug feature representations.

| Combination of disease and drug representation ($\mathcal{V}.\mathcal{U}$) | $e = ACC$ | $e = AUC$ | $e = AUC - PR$ | $e = WAS$ |
|---|---|---|---|---|
| $B^T.W$ | %79.4 | %71.4 | %53.4 | %65.8 |
| $B^T.O$ | %80.5 | %74.0 | %56.8 | %68.4 |
| $C^T.W$ | **%82.0** | %75.6 | **%59.1** | **%70.3** |
| $C^T.O$ | %79.9 | %72.9 | %55.4 | %67.3 |
| $B^D.W$ | %80.2 | %72.6 | %54.9 | %67.0 |
| $B^D.O$ | %81.2 | **%75.9** | %53.6 | %68.0 |
| $C^D.W$ | %79.0 | %75.5 | %58.2 | %69.3 |
| $C^D.O$ | %78.2 | %73.1 | %54.7 | %66.8 |
| $B^S.W$ | %80.9 | %72.6 | %56.8 | %67.9 |
| $B^S.O$ | %77.9 | %67.3 | %49.2 | %62.2 |
| $C^S.W$ | %67.1 | %60.5 | %38.5 | %53.0 |
| $C^S.O$ | %72.4 | %63.0 | %41.3 | %56.2 |
| $B^C.W$ | %76.9 | %69.3 | %51.1 | %63.5 |
| $B^C.O$ | %75.5 | %66.9 | %48.0 | %61.0 |
| $C^C.W$ | %77.0 | %72.2 | %55.2 | %66.4 |
| $C^C.O$ | %75.9 | %69.0 | %49.5 | %62.6 |
| $B^G.W$ | %78.5 | %68.9 | %51.1 | %63.7 |
| $B^G.O$ | %76.4 | %65.5 | %46.4 | %60.0 |
| $C^G.W$ | %77.6 | %70.4 | %53.3 | %65.0 |
| $C^G.O$ | %76.1 | %67.4 | %49.0 | %61.8 |
| $C^N.W$ | %70.5 | %58.3 | %37.9 | %52.6 |
| $C^N.O$ | %77.2 | %69.0 | %49.4 | %62.8 |

- **Assessment of the effectiveness of classification methods**

Here, every classifier ($\mathcal{C} \in \mathbb{C}$) uses a voting ensemble training approach ($\mathcal{T} \in \mathbb{T}$) for learning. We analyze all combinations of classifiers and training approaches to declare the best model owing to (16). The results are shown in table 8.

$$avg(\mathcal{C}^z = \mathcal{C}, \mathcal{T}^z = \mathcal{T}) = \frac{1}{|\mathbb{V}|.|\mathbb{U}|} \sum_{\mathcal{V} \in \mathbb{V}} \sum_{\mathcal{U} \in \mathbb{U}} e(model^z) \qquad (16)$$

The ACC criterion rises if the number of negatives grows. So, for every three classifiers, the ACC on $T_5$ is better than the others. On the other hand, by increasing the negative samples, the amount of AUC-PR is decreased. We define WAS score to make a better trade-off among ACC, AUC and AUC-PR.

According to corresponding results, our model using decision tree as a classifier and $T_1$ as a training approach is performed better than other ones.

Table8: Evaluation criteria based on classification method

| Classifier ($\mathcal{C}$) | Training approach ($\mathcal{T}$) | $e = Acc$ | $e = AUC$ | $e = AUC - PR$ | $e = WAS$ |
|---|---|---|---|---|---|
| DT | $T_1$ | %77.7 | %77.9 | **%72.3** | **%75.6** |
|  | $T_2$ | %81.1 | %76.9 | %62.8 | %72.1 |
|  | $T_3$ | %87.4 | **%79.0** | %62.2 | %74.0 |
|  | $T_5$ | **%89.6** | %73.9 | %49.1 | %67.1 |
| RF | $T_1$ | %72.1 | %72.3 | %67.3 | %70.2 |
|  | $T_2$ | %77.3 | %69.9 | %53.6 | %64.8 |
|  | $T_3$ | %84.0 | %71.5 | %52.0 | %66.2 |
|  | $T_5$ | %87.6 | %66.4 | %38.4 | %59.4 |
| CNB | $T_1$ | %63.6 | %63.6 | %59.1 | %61.8 |
|  | $T_2$ | 0.6716 | 0.6185 | 0.4172 | 0.5486 |
|  | $T_3$ | 0.6799 | 0.5963 | 0.3093 | 0.4982 |
|  | $T_5$ | 0.7187 | 0.6234 | 0.2314 | 0.4857 |

- **Comparing DRP-VEM with DisDrugPred**

We implement DisDrugPred [10] and analyze its performance by utilizing our dataset. As DisDrugPred is a regression algorithm and not a classifier, we calculate its mean square error (MSE) instead of the ACC. As mentioned DisDrugPred article, we perform 5-fold cross-validation.

According to our assessment, the accurate combination of parameters belongs to model $<\mathcal{V}=W, \mathcal{U}=C^T, \mathcal{C}=DT, \mathcal{T}=T_1>$, that we named *BestOverAll*. In the *BestOverAll*, the Wang vector representation for disease ($W$) and target cosine similarity vector for drug ($C^T$) are combined to fed $DT$ classifier, which is learned by performing voting ensemble training $T_1$. This model is generally preferred. However, the best scores among all 264 experimented models belong to model *BestAmongAll* $=<\mathcal{V}=W, \mathcal{U}=B^S, \mathcal{C}=DT, \mathcal{T}=T_3>$, where the combination of Wang vector for disease ($W$) and side effect binary vector for drug ($B^S$) is fed to classifier $DT$ learned based on one-to-three distribution training set $T_3$.

As the ACC score is not available for DisDrugPred and similarly MSE is not available for *BestOverAll* and *BestAmongAll*, we calculate WAS score as the average of AUC and AUC-PR. Table 9 illustrates the corresponding results. *BestOverAll* is slightly better than DisDrugPred. But *BestAmongAll* achieves remarkable evaluation scores.

*Table9:Evaluation criteria based on classification method.*

| Model | MSE | ACC | AUC | AUC-PR | WAS |
|---|---|---|---|---|---|
| **DisDrugPred** | 0.01191 | - | %82.4 | %63.9 | %73.1 |
| *BestAmongAll* | - | %97.7 | %97.2 | %92.1 | %94.6 |
| *BestOverAll* | - | %81.9 | %81.8 | %76.6 | %79.2 |

# Conclusions

This article proposed a new framework, named DPR-VEM, to solve the DR problem using a voting ensemble method. We examined different parameters to find the proper combination of drug and disease feature representations, classification method and training approach. We chose ACC, AUC, AUC-PR and WAS to evaluate the framework. Owing to results, the best overall model (*BestOverAll*) belongs to target cosine similarity vector as drug feature representation, Wang similarity vector as disease representation decision tree as classification method, and one-to-one distribution as voting-ensemble training approach. The ACC, AUC, AUC-PR and WAS scores for the *BestOverAll* model are %81.9, %81.8, %76.6 and %79.7, respectively. DRP-VEM is compared with DisDrugPred [10] as a state-o-the-art drug repositioning method. DisDrugPred got AUC = %82.4, AUC-PR = %63.9 and WAS = %73.1.

The data and implementation of DRP-VEM is available at http://bioinformatics.aut.ac.ir/DRP-VEM.

In conclusion, using target or domain as a drug feature is necessary, while concatenating all features reduces the model's accuracy and caused redundancy. The performance of the cosine similarity vector and Wang vector as drug and disease feature representation is more accurate. Moreover, the decision tree classifier distinguishes the dataset better than others. In addition, applying voting ensemble approach to make training and test sets solve the classification method's biasing challenge.

In this article, we focused more on the assessment of drug feature representation and less on disease. In the future, we aim to analyze disease feature representations more. This study, utilized fingerprint as a drug chemical structure representation. While SMILES representation appeared more informatic, we want to achieve a representation format for SMILES.

# Declarations

## Acknowledgment

We would like to express our great appreciation to Miss. Mina Shaygan, the member of the CBRC lab at Amirkabir university of technology, for her patience in designing the CBRC lab webpage. Her willingness to give her time so generously is very much appreciated.

## Authors' contributions

FZ and ZG contributed to the design and implementation of the framework, the analysis of the results, and the manuscript's writing. BM contributed to implementing DisDrugPred approach.

## Funding

No funding information to declare.

## Availability of data and materials

The data that support the findings of this study, including disease similarity based on Wang's method, drug name, identifiers and targets, drug domain, drug side effects, drug chemical structure, and drug-disease associations are openly available; namely DincRNA [18], DrugBank [14], Uniprot [15], SIDER4.1[16], PubChem [17], and repoDB [13] have been downloaded and used in this study, respectively.

## Ethics approval and consent to participate

Not applicable.

## Consent to publication

All authors give consent to publish.

## Competing interests

No competing interest to declare.

## Author Details

1,2,3 Department of Mathematics and Computer Science, Amirkabir University of Technology, Tehran, Iran.

# Resources